\documentclass[twocolumn,showpacs,showkeys]{revtex4}
\usepackage{graphicx}
\usepackage{bm}
\usepackage{color}
\usepackage{amsmath}
\usepackage{natbib}

\begin{document}

\title{Electrostatic Langmuir and SEAWs in spin polarized plasma double layer}

\author{Pavel A. Andreev}
\email{andreevpa@physics.msu.ru}
\affiliation{Department of General Physics, Faculty of physics, Lomonosov Moscow State University, Moscow, Russian Federation.}

\author{T. G. Kiriltseva}
\email{kiriltsevatg@rambler.ru}
\affiliation{Department of General Physics, Faculty of physics, Lomonosov Moscow State University, Moscow, Russian Federation.}

\author{Punit Kumar}
\email{punitkumar@hotmail.com}
\affiliation{Department of Physics University of Lucknow, Lucknow-226007, India.}

\author{L. S. Kuz'menkov}%
\email{lsk@phys.msu.ru}
\affiliation{Department of Theoretical Physics, Faculty of physics, Lomonosov Moscow State University, Moscow, Russian Federation.}

\date{\today}

\begin{abstract}
The quantum hydrodynamic model of the electrostatic waves in the two parallel layers of two dimensional electron gases (2DEGs) is developed.
It is considered for two regimes: classic regime and quantum regime with the separate spin evolution.
Two Langmuir-like waves are found in classic case
which have an interference-like pattern in the frequency dependence on concentration $\omega^{2}\sim(n_{01}+n_{02}\pm2\sqrt{n_{01}n_{02}})$.
They appear instead of two 2D Langmuir waves in noninteracting 2DEGs.
The spectrum of four waves is found in the quantum regime.
Two extra waves are related to the separate spin evolution and associated to the spin-electron acoustic waves.
The influence of the quantum Bohm potential is considered either.
\end{abstract}




\maketitle


\section{Introduction}

Bulk plasmas phenomena, surface plasma phenomena, and phenomena in two dimensional electron gas are intensively studied.
It is possible to create several layers of the two dimensional electron gas in one sample.
Therefore, this paper is focused on the wave phenomena in two layers of the two dimensional electron gas (2DEG) \cite{Bounouh IEEE 03}.
One layer of the two dimensional electron gas demonstrates the Langmuir wave existence.
The ion-sound also exists in 2DEG at the account of the ion motion,
but the high-frequency waves are under consideration in this work.
The double 2DEG (double-quantum-well) has more complex spectrum of collective excitations.
The Langmuir waves in each layer show interference-like picture.
First of all this interference is in the frequency dependence
on the partial concentrations of each layer.

Compositions of several 2DEGs or 2DEGs and two-dimensional hole gases are applied in modern devises,
for instance they are utilized in the field effect transistors \cite{Schubert IEEE 86, Chakhnakia Proc. SPIE 04, Lin IEEE 03, Aleksov SST 03}.
Presence of the p-type quantum well produces a two-dimensional gas of holes \cite{Nakagawa APL 89}.
Different kinds of double wells are used in the matter-wave interferometry \cite{Schumm NP 05}.
Let us to point out that methods of creation of high quality double 2DEGs are developed many years ago \cite{Chui APL 95}.
The spin effects in 2DEGs and double 2DEGs are studied experimentally.
For instance, the spin coherence is found in the regime of high-mobility 2DEGs \cite{Ullah JAP 16}.

The double 2DEG is interesting from several points of view.
If potential barrier for the carriers is relatively small there is the tunneling effect.
Corresponding exchange of electrons between layers modifies the initial concentrations of electrons in 2DEGs.
Such system shows a similarity to the two level system
which has been studied in the atomic systems,
where Landau-Zener transitions and many other effects have been observed.
The double-well potential trap with the ultracold atoms demonstrates similarity
as well \cite{Nesterenko JP B 09}, \cite{Nesterenko JP B 12}, \cite{Nesterenko LP 14}.
However, we are focused on the regime of the large potential barrier.
The time scales are restricted by Landau-Zener transitions effect related to the small tunneling.
Hence, the relatively fast effects are studied.
Therefore, we consider the high frequency perturbations of the equilibrium state.


Moreover, we are interested in the influence of the spin effects on the properties of waves in several interacting degenerate 2DEGs.
To be more specific, we are focused on the separate spin evolution (SSE) contribution.
As it is well-known, the SSE requires description of electrons as two species system
since spin-up electrons and the spin-down electrons are considered as two different species \cite{Andreev PRE 15}.
As a natural consequence,
the spin electron acoustic wave (SEAW)
or the spin plasmons appears in plasmas \cite{Andreev PRE 15}, \cite{Andreev EPL 16}, \cite{Ryan PRB 91}, \cite{Agarwal PRB 14}.
Since derivation of the quantum hydrodynamic equation for collection of interacting charged spin-1/2 particles in 2001
the electrons considered in traditional way as the single fluid with additional characteristic:
the spin density or the spin caused magnetization \cite{MaksimovTMP 2001}.
It was suggested in 2004 that electrons in quantum plasmas can be considered as two different fluids for two different spin projections \cite{Harabadze RPJ 04}.
Such description is used in some later papers (see for instance \cite{Brodin PRL 08 Cl Reg}, \cite{Brodin PRL 10 SPF}).
However, a rigorous derivation of the hydrodynamic model based on this concept showed that part of the intuitively introduced variables have no physical meaning \cite{Andreev PRE 15}.
Besides several essential parts of the model were lost, such as the difference of the Fermi pressure in different subspecies of electrons
and unconservation of particles in each subspecies at the conservation of the total number of electrons \cite{Andreev PRE 15}.

Mostly, the bulk SEAWs are studied in the plasmas
\cite{Andreev PRE 15}, \cite{Andreev AoP 15}, \cite{Andreev PoP 18 Extr SEAW Bohm}, \cite{Andreev PoP 18 kin obl}, \cite{Iqbal PLA 18}.
Besides, the surface SEAWs \cite{Andreev APL 16} and the SEAWs in 2DEGs \cite{Andreev EPL 16} are studied.
Therefore, the study of the SEAWs in the multi-layered structures is the direct consequence of Ref. \cite{Andreev EPL 16}.
The SEAWs exist among different spin effects in quantum plasmas \cite{Takabayasi PTP 55 a}, \cite{Marklund PRL07}, \cite{Andreev PoP 17 kin I}, \cite{Andreev PoP 17 kin II}, \cite{Andreev PoP 17 kin Non Triv}.

Influence of the spin effects on the dielectric permeability of plasmas at the arbitrary temperatures is considered in Ref. \cite{Bobrov PP 18}.
It is found that the plasma can demonstrate the paramagnetic behavior.

The wave phenomena affects different processes in the plasma-like substances, properties of turbulence, the heat transfer, etc.
Therefore, the knowledge of the wave spectra in double spin-polarized 2DEG is a step towards understanding of the future role of the spin-polarized 2DEGs in construction of the field effect transistors.

This paper is organized as follows.
In Sec. II the classic linear electrostatic waves in two interacting 2DEGs are considered via the hydrodynamic method.
In Sec. III the influence of the separate spin evolution and properties of the SEAWs electrostatic waves in two interacting 2DEGs are studied.
In Sec. IV a brief summary of obtained results is presented.

\section{waves in double 2DEG: classic regime}

\subsection{Hydrodynamic equations in the electrostatic limit}

Hydrodynamic equations for the two dimensional objects can be written in the following form \cite{Fetter AnP 73}, \cite{Fetter Anp 74}
\begin{equation}\label{SUSD2D cont eq electrons spin s}
\partial_{t}n_{ej}+\nabla\cdot(n_{ej}\textbf{v}_{ej})=0, \end{equation}
and
$$mn_{ej}(\partial_{t}+\textbf{v}_{ej}\cdot\nabla)\textbf{v}_{ej}+\nabla p_{ej}$$
\begin{equation}\label{SUSD_2L2D}=q_{e}n_{ej}\biggl(-q_{e}\nabla\int \frac{n_{e1}+n_{e2}-n_{0}}{\mid \textbf{r}-\textbf{r}'\mid}d\textbf{r}'+\frac{1}{c}[\textbf{v}_{ej},\textbf{B}_{ext}]\biggr),\end{equation}
where $j=1,2$ for different layers, $n_{0}$ presents the contribution of motionless ions.
We note that $\textbf{v}_{j}=\{v_{jx},v_{jy}\}$ and $n_{j}=n_{j}(x,y)$, $\textbf{v}_{j}=\textbf{v}_{j}(x,y)$, $[n_{j}]=$cm$^{-2}$ for plane-like 2DEG.

Let us focus on the integral term since it is written in rather simplified form.
We rewrite it in terms of three dimensional variables
$$\textbf{E}(\textbf{R},t)= -q_{e}\nabla$$
$$\int \frac{n_{e1}\delta(z')+n_{e2}\delta(z'-d)-n_{0i,1}\delta(z')-n_{0i,2}\delta(z'-d)}{\mid \textbf{R}-\textbf{R}'\mid}d\textbf{R}',$$
where $\textbf{R}=\{\textbf{r},z\}=\{x,y,z\}$.

We are focused on the calculation of the linear wave spectrum.
We consider the plane waves
(which wave front actually is a line,
but it shows similarity to the plane waves in three-dimensional mediums)
propagating in the $x$-direction.
Hence, the perturbations can be written as follows
$\delta n_{j}=N_{j} e^{-\imath\omega t+\imath \textbf{k}\textbf{r}}$
and $\delta \textbf{v}_{j}=\textbf{V}_{j} e^{-\imath\omega t+\imath \textbf{k}\textbf{r}}$,
where $\textbf{k}=\{k_{x},0,0\}$.
Moreover, the full concentration and the velocity field traditionally appear as the superposition of the equilibrium values
and the perturbations $n_{j}=n_{0j}+\delta n_{j}$ and $\textbf{v}_{j}=0+\delta \textbf{v}_{j}$.

The Fermi pressure is used for the equation of state: $p_{ej}=\pi\hbar^{2}n_{ej}^{2}/2m$.
Presence of the second layer modifies the equation of state \cite{Kurobe PRB 94},
but we do not include this effect here.

Next, we present linearized and Fourier transformed equations
\begin{equation}\label{SUSD2D cont eq linearised}
-\imath\omega\delta n_{ej}+\imath k_{x}n_{0ej}\delta v_{ej,x}=0, \end{equation}
$$mn_{0ej}(-\imath\omega)\delta v_{ej,x}+\imath k_{x} \frac{\partial p_{ej}}{\partial n_{ej}}\delta n_{ej}$$
\begin{equation}\label{SUSD_2L2D Euler x linearised} =q_{e}n_{0ej}\biggl(-\frac{2\pi q_{e}}{k}\imath k_{x}(\delta n_{ej}+ e^{-kd}\delta n_{ej'}) +\frac{1}{c}v_{ej,y}B_{0}\biggr),\end{equation}
and
\begin{equation}\label{SUSD_2L2D Euler y linearised} mn_{0ej}(-\imath\omega)v_{ej,y} =q_{e}n_{0ej}\frac{1}{c}v_{ej,x}B_{0}. \end{equation}
The projection of the wave vector has the following form in the considering regime $k_x=\pm k$,
where the sign depends on the direction of wave propagation.
Without loss of generality, we assume $k_x=k$.
Since, final equations contain $k_x^2$
which does not depend on the wave propagation direction.

It is assumed that the potential barrier between two layers is sufficiently large.
Otherwise, there is exchange of electrons between 2DEGs.
It would create extra terms in the continuity and Euler equations responsible for the transitions of electrons
and number of electrons in each layer would not conserve.
The model would be similar to the one presented below for the separate spin evolution,
where each spin state in the fixed layer can be identified with the one of two layers.

Here, we have calculated the integral term representing the electric field.
The result for the electric field created by plane located at $z=0$ is
$\delta\textbf{E}(\omega,\textbf{k},z)=2\pi\imath q_{e}\textbf{k}\delta n_{e1} e^{-kz}/k$.
Hence, it acts on the plane located at $z=0$ (the selfaction of the plane) we find traditional for the 2DEG result $\delta\textbf{E}(\omega,\textbf{k},0)=2\pi\imath q_{e}\textbf{k}\delta n_{e1}/k$.
If we consider its action on the another plane which is located at $z=d$
we obtain that the electric field contribution is reduced by factor $e^{-kd}$.
Similar result is found for the electric field created by the plane located at $z=d$:
$\delta\textbf{E}(\omega,\textbf{k},z)=2\pi\imath q_{e}\textbf{k}\delta n_{e2} e^{-k(z-d)}/k$.

The transport of particles between two layers exists
if there is no restriction on the potential barrier.
Hence, the probability to find each particle in chosen layer changes.
However, this probability (the corresponding wave function) can be used for the construction of the concentration.
The full many-particle wave function gives three dimensional dynamics of each electron.
However, if we consider a time scale large in compare with time of the transition of the electron $t\gg\tau$,
we can restrict the analysis by the description of two interacting layers.
Hence, the present hydrodynamic model is applicable for the neglegible transport or for the nontrivial transport,
but for the relatively large time scale.


\subsection{The classic wave solutions}

The linearized and Fourier transformed hydrodynamic
equations (\ref{SUSD2D cont eq linearised})-(\ref{SUSD_2L2D Euler y linearised})
are the algebraic equations.
Moreover, this is a homogeneous set of equations.
Existence of waves requires existence of the nonzero perturbations $\delta n_{j}$ and $\delta \textbf{v}_{j}$.
Corresponding nontrivial solution exists if the determinant of this set of equations is equal to zero.
This condition provides an equation
which defines the spectrum.
However, in order to simplify the calculations we substitute $\delta v_{jy}$ from equation (\ref{SUSD_2L2D Euler y linearised})
to equation (\ref{SUSD_2L2D Euler x linearised}) and the perturbation of concentration from equation (\ref{SUSD2D cont eq linearised}) to equation (\ref{SUSD_2L2D Euler x linearised}).
At this point find
$\delta n_{j}=n_{0ej} \imath k_x \delta v_{jx}/\omega$ and
$\delta v_{jy}=-\imath\Omega\delta v_{jx}/\omega$.

Hence, the set of six equations (\ref{SUSD2D cont eq linearised})-(\ref{SUSD_2L2D Euler y linearised})  simplifies to two equations
\begin{equation}\label{SUSD_2L2D } (\omega^{2}-k_{x}^{2}U_{1}^{2}-\Omega^{2})\delta n_{1}=\omega_{L1}^{2}(\delta n_{1}+e^{-kd}\delta n_{2}), \end{equation}
and
\begin{equation}\label{SUSD_2L2D } (\omega^{2}-k_{x}^{2}U_{2}^{2}-\Omega^{2})\delta n_{2}=\omega_{L2}^{2}(\delta n_{2}+e^{-kd}\delta n_{1}), \end{equation}
where $\omega_{Lj}^{2}=2\pi q^{2}n_{0ej}k/m$ and $U_{j}^{2}=\pi\hbar^{2}n_{0ej}/m^{2}=v_{Fe,2D}^{2}/2$.

Consequently, the determinant of this set of equations is a two on two determinant
which can be easily calculated.
As a result we find the dispersion equation of the two layers spectrum
$$(\omega^{2}-\Omega^{2}-\omega_{L1}^{2}-k^2 U_{1}^2)(\omega^{2}-\Omega^{2}-\omega_{L2}^{2}-k^2 U_{2}^2) $$
\begin{equation}\label{SUSD_2L2D classic disp eq} -e^{-2kd}\omega_{L1}^{2}\omega_{L2}^{2}=0. \end{equation}


Equation (\ref{SUSD_2L2D classic disp eq}) is the quadratic equation relatively $\omega^{2}$.
Therefore, its solutions can be written in the analytical form as follows:
$$\omega^{2}=\frac{1}{2}\Biggl(\sum_{j=1}^{2}\omega_{Lj}^{2}+2\Omega^{2}+k_{x}^{2}(U_{1}^{2}+U_{2}^{2})$$
\begin{equation}\label{SUSD_2L2D classic spectrum}
\pm\sqrt{[\omega_{L1}^{2}-\omega_{L2}^{2}+k_{x}^{2}(U_{1}^{2}-U_{2}^{2})]^{2}+4\omega_{L1}^{2}\omega_{L2}^{2}e^{-2kd}}\biggr). \end{equation}

\begin{figure}
\includegraphics[width=8cm,angle=0]{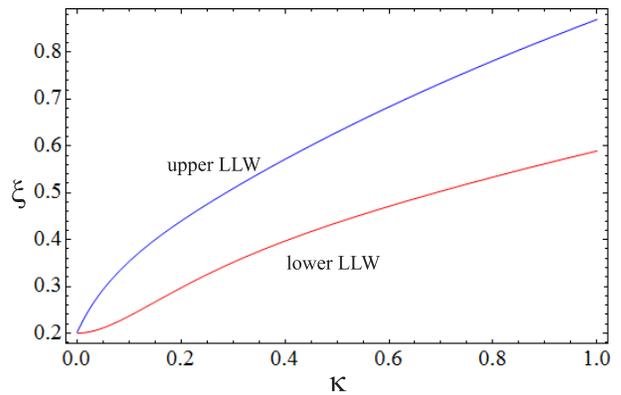}
\caption{\label{SUSD ObEx 01}
Solutions of equation (\ref{SUSD_2L2D classic spectrum}) are presented.
It demonstrates a spectrum of two waves.
Here we use the following dimensionless notations
$\xi=\omega/\omega_{L0}$, $b=\Omega/\omega_{L0}$, $\kappa=k/\sqrt{n_{0}}$, where $n_{0}=n_{01}+n_{01}$, and $\omega_{L0}^{2}=2\pi e^{2}n_{0}^{3/2}/m$.
Moreover, we use the dimensionless partial concentrations $c_{j}=n_{0j}/n_{0}$, the dimensionless interlayer distance $q=d\sqrt{n_{0}}$ and
the dimensionless Bohr radius $r\equiv r_{B}\sqrt{n_{0}}$, where $r_{B}=\hbar^{2}/me^{2}$.
Parameter $r$ fixes the full concentration of electrons in the system.
Here and below we use $b=0.2$.
Here we also have $q=10$, $r=0.009$, $c_{1}=0.3$ and $c_{2}=0.7$.}
\end{figure}

Presence of the second layer shows similarity to a single layer containing two species of charged particles.
It reveals in appearance of an extra wave.
Usually, it is the sound-like wave.
The sound wave appear both for the 2D systems and the bulk systems.
Simplest example is the ion-sound existing at the description of the electron-ion systems.
There are positron-acoustic wave in the electron-positron
and electron-positron-ion plasmas \cite{Nejoh AJP 96}, \cite{Tsintsadze EPJ D 11}, \cite{Andreev_Iqbal PoP 16}.
It appears as the result of the dynamical properties of composed systems of charged particles.
A comprehensive analysis of the electrostatic perturbation in composed systems is presented in Ref. \cite{Rightley 1807}.
The SEAW is the another example,
where electrons with different spin projections behave as two different species \cite{Andreev PRE 15}, \cite{Andreev EPL 16}.
However, in this case the spectrum is different.
The Coulomb nature of both branches dominates over the pressure
effects at the formation of the second wave and
we have two Langmuir-like waves (Llws) with the interference picture having opposite signs of the interference terms (see below the simplified spectrum).

The double and triple 2DEGs are studied in Ref. \cite{Fertig PRB 89},
where the spectra of the collective excitations are found.
Ref. \cite{Fertig PRB 89} applies the second quantization formalism in the self-consistent approximation
with further reference on the results of the random-phase approximation \cite{Paquet PRB 85}.
The occupation of the lowest Landau level in the strong magnetic field limit is assumed in Ref. \cite{Fertig PRB 89}.
Moreover, the wave functions are constructed via operators of creation of the electron-hole pairs.
Hence, the exitons dynamics is assumed there (see also \cite{Chen PRL 91}).
It shows that a different physical picture is considered in Refs. \cite{Fertig PRB 89}, \cite{Chen PRL 91}.
Moreover, they have different spectrum for the double 2DEG either.
The single curve is found in Ref. \cite{Fertig PRB 89}.
The spectrum is related to the neutrality of the considered electron-hole pairs,
while our analysis assumes a possibility of local violation of the electric neutrality during electron oscillations.

Consider a regime of zero magnetic field.
Moreover, consider a regime, where concentrations in two layers have small difference $\mid n_{01}-n_{01}\mid\ll n_{0j}$
and the distance between layers is relatively small.
Hence, the exponent in the last term under the square root is of order of $e^{-2kd}\sim e^{-1}$.
Therefore, the last term under the square root dominates over the other terms located under the square root.
It provides the following spectrum
\begin{equation}\label{SUSD_2L2D classic spectrum limit 1} \omega^{2}=\frac{1}{2}(\omega_{L1}^{2}+\omega_{L2}^{2}) +\frac{1}{2}k^{2}(U_{1}^{2}+U_{2}^{2})\pm \omega_{L1}\omega_{L2} e^{-kd}. \end{equation}
Notice the functional structure of the dependence on the wave vector $k$: $\omega^{2}=ak+bk^2\pm cke^{-kd}$.
For the small wave vectors $k\rightarrow0$,
we have $e^{-kd}\sim 1$,
hence $\omega^{2}=\frac{1}{2}(\omega_{L1}\pm\omega_{L2})^{2}$.
Consequently, the frequency square $\omega^{2}$ is proportional to the $(n_{01}+n_{02}\pm2\sqrt{n_{01}n_{02}})k$.

In this regime we have two square-root spectra.
It mirrors the 2D Langmuir wave spectra,
but we have the interference-like pattern in the concentration dependence.

The presence of the magnetic field (at the neglecting of the spin effects) leads to the shift of $\omega^{2}$ on $\Omega^{2}$.

Numerical solution of equation (\ref{SUSD_2L2D classic spectrum}) is given by Fig. (\ref{SUSD ObEx 01a}),
where the following equilibrium concentrations in two 2DEGs $n_{01}=10^{10}$ cm$^{-2}$ and $n_{02}=2.33\times10^{10}$ cm$^{-2}$.
It leads to the following value of dimensionless Bohr radius $r=\sqrt{n_{0}}r_{B}=9.2\times10^{-4}$,
where $n_{0}=n_{01}+n_{02}$.
The distance between layers is $d=0.55\times10^{-4}$ cm.

\section{regime of the spin influence}

\subsection{SSE hydrodynamic equations}

Next, consider an advanced regime, where the electron gas in each layer is the partially spin polarized gas.
The SSE-QHD can be applied to the electrostatic waves in this object.
These equations are derived in \cite{Andreev PRE 15}.
They were further adopted for the two dimensional systems in \cite{Andreev EPL 16}.
Finally, we have the following equations,
where electrons with a chosen spin projection in each layer are considered as an independent species.

Corresponding four continuity equations can be written in compact form via the application of subindexes $j$ and $s$:
\begin{equation}\label{SUSD2D cont eq electrons spin s}
\partial_{t}n_{js}+\nabla(n_{js}\textbf{v}_{js})=(-1)^{i_{s}}T_{jz}, \end{equation}
where $s=\{u=\uparrow, d=\downarrow\}$ is the subindex describing the spin state of subspecies of electrons,
$T_{jz}=\frac{\gamma_{e}}{\hbar}(B_{x}S_{jy}-B_{y}S_{jx})$ is the z-projection of torque presented in Cartesian coordinates,
$i_{s}$: $i_{u}=2$, $i_{d}=1$,
with the spin density projections $S_{jx}$ and $S_{jy}$,
each of them is a mix of $\psi_{ju}$ and $\psi_{jd}$
which are components of the wave spinor.
The explicit forms of $S_{jx}$ and $S_{jy}$ appear as
$S_{jx}=\psi_{j}^{*}\sigma_{x}\psi_{j} =\psi_{jd}^{*}\psi_{ju}+\psi_{ju}^{*}\psi_{jd} =2a_{ju}a_{jd}\cos\Delta \phi_{j}$, $S_{jy}=\psi_{j}^{*}\sigma_{y}\psi_{j}=\imath(\psi_{jd}^{*}\psi_{ju}-\psi_{ju}^{*}\psi_{jd})=-2a_{ju}a_{jd}\sin\Delta \phi_{j}$,
where $\Delta \phi_{j}=\phi_{ju}-\phi_{jd}$,
$\psi_{j}$ is the wave spinor composed of $\psi_{ju}$ and $\psi_{jd}$,
$\phi_{js}$ and $a_{js}$ are the phases and the amplitudes of the partial wave functions.
Spin projections are not related to different species of electrons with different spin directions.
Functions $S_{jx}$ and $S_{jy}$ describe the simultaneous evolution of both species.

The four vector Euler equations can be written in the following form
$$mn_{js}(\partial_{t}+\textbf{v}_{js}\cdot\nabla)\textbf{v}_{js}+\nabla p_{js}-\frac{\hbar^{2}}{2m}n_{js}\nabla\Biggl(\frac{\triangle \sqrt{n_{js}}}{\sqrt{n_{js}}}\Biggr)$$
$$=q_{e}n_{js}\biggl(-q_{e}\nabla\int \frac{n_{1u}+n_{1d}+n_{2u}+n_{2d}-n_{0}}{\mid \textbf{r}-\textbf{r}'\mid}d\textbf{r}' $$ $$+\frac{1}{c}[\textbf{v}_{js},\textbf{B}]\biggr)+(-1)^{i_{s}}\gamma_{e}n_{js}\nabla B_{z}$$
\begin{equation}\label{SUSD2D Euler eq electrons spin UP} +\frac{\gamma_{e}}{2}(S_{jx}\nabla B_{x}+S_{jy}\nabla B_{y})+(-1)^{i_{s}}m(\widetilde{\textbf{T}}_{jz}-\textbf{v}_{js}T_{jz}),\end{equation}
with $\widetilde{\textbf{T}}_{jz}=\frac{\gamma_{e}}{\hbar}(\textbf{J}_{j(M)x}B_{y}-\textbf{J}_{j(M)y}B_{x})$,
which is the torque current,
where
\begin{equation}\label{SUSD2D Spin current x} \textbf{J}_{j(M)x}=\frac{1}{2}(\textbf{v}_{ju}+\textbf{v}_{jd})S_{jx}
-\frac{\hbar}{4m} \biggl(\frac{\nabla n_{ju}}{n_{ju}}-\frac{\nabla n_{jd}}{n_{jd}}\biggr)S_{jy}, \end{equation}
and
\begin{equation}\label{SUSD2D Spin current y} \textbf{J}_{j(M)y}= \frac{1}{2}(\textbf{v}_{ju}+\textbf{v}_{jd})S_{jy}
+\frac{\hbar}{4m}\biggl(\frac{\nabla n_{ju}}{n_{ju}}-\frac{\nabla n_{jd}}{n_{jd}}\biggr)S_{jx}, \end{equation}
where $q_{e}=-e$, $\gamma_{e}=-\frac{e\hbar}{2mc}$ is the gyromagnetic ratio for electrons, $p_{sj}=\pi\hbar^{2}n_{sj}^{2}/m$.
Functions $\textbf{J}_{j(M)x}$ and $\textbf{J}_{j(M)y}$ are elements of the spin current tensor $J^{\alpha\beta}_{j}$.

\begin{figure}
\includegraphics[width=8cm,angle=0]{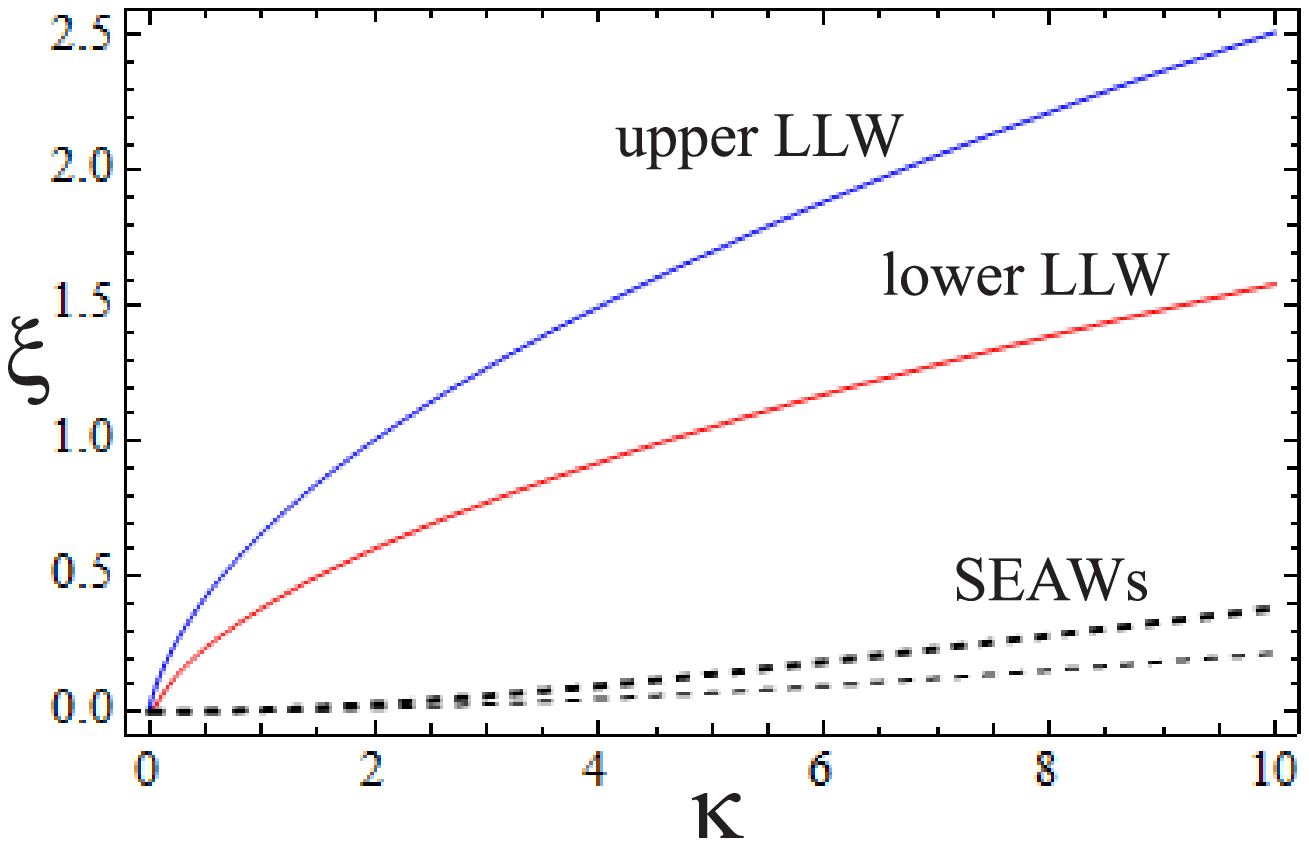}
\caption{\label{SUSD ObEx 02}
Solutions of equation (\ref{SUSD_2L2D GENERAL spectrum}) are demonstrated at the following parameters
$c_{1}=0.3$, $\eta_{1}= 0.1$, $c_{2}=0.7$, $\eta_{2}=0.2$, $b=0.2$, $r=0.009$, $q = 10$ (no quantum Bohm potential is included).
These $c_{i}$ correspond to $n_{01}=10^{12}$ cm$^{-2}$ and $n_{02}=2.33\times10^{12}$ cm$^{-2}$.}
\end{figure}

\begin{figure}
\includegraphics[width=8cm,angle=0]{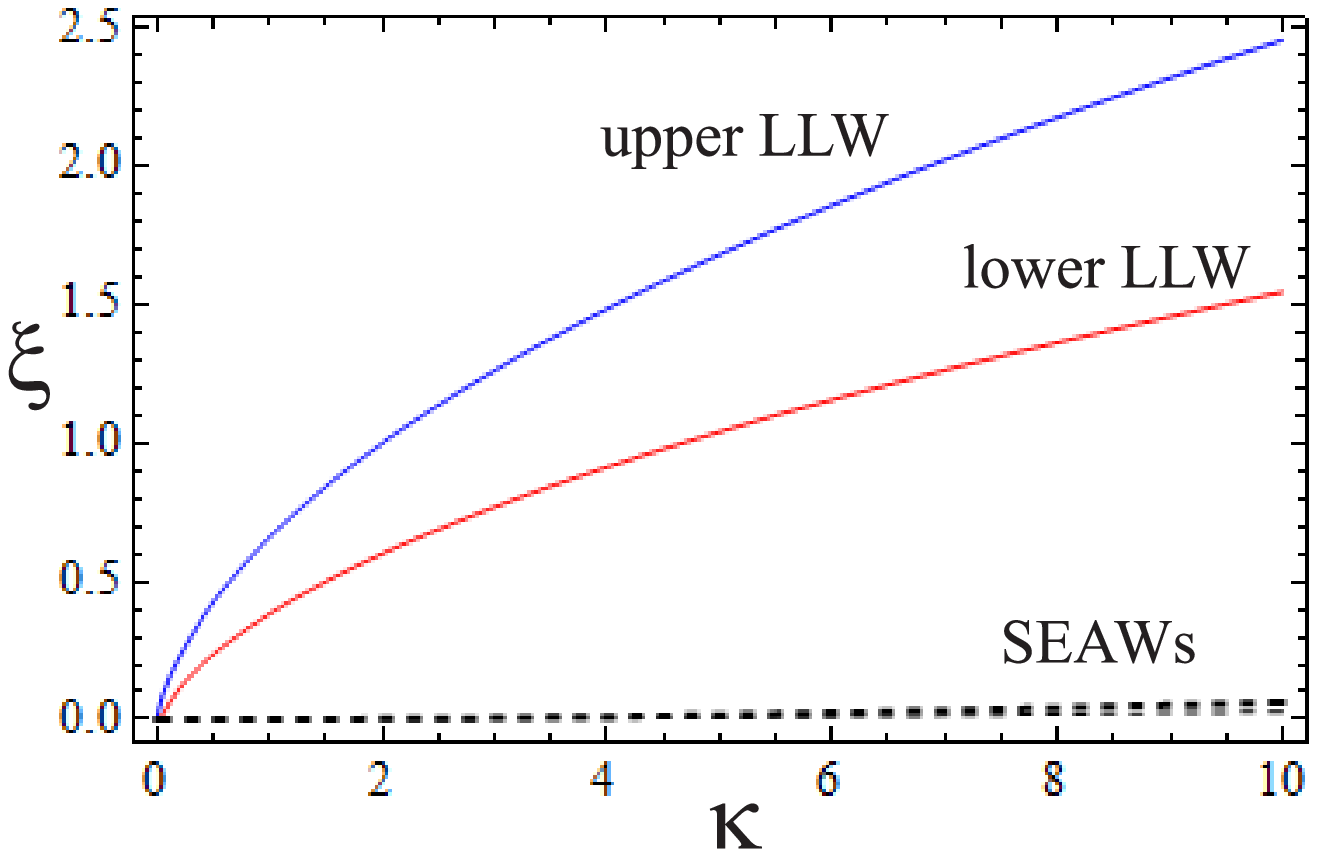}
\caption{\label{SUSD ObEx 03}
Solutions of equation (\ref{SUSD_2L2D GENERAL spectrum}) are demonstrated at the following parameters
$c_{1}=0.3$, $\eta_{1}= 0.1$, $c_{2}=0.7$, $\eta_{2}=0.2$, $b=0.2$, $r=0.0009$, $q = 10$ (no quantum Bohm potential is included).
These $c_{i}$ correspond to $n_{01}=10^{10}$ cm$^{-2}$ and $n_{02}=2.33\times10^{10}$ cm$^{-2}$.}
\end{figure}

\begin{figure}
\includegraphics[width=8cm,angle=0]{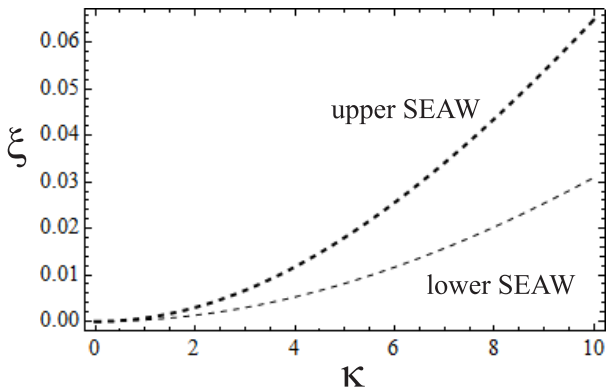}
\caption{\label{SUSD ObEx 04}
The low-frequency solutions of equation (\ref{SUSD_2L2D GENERAL spectrum}) are shown (no quantum Bohm potential is included).
This is the low frequency part of Fig. (\ref{SUSD ObEx 03})
The pair of SEAWs forms the low-frequency part of the spectrum.}
\end{figure}

\begin{figure}
\includegraphics[width=8cm,angle=0]{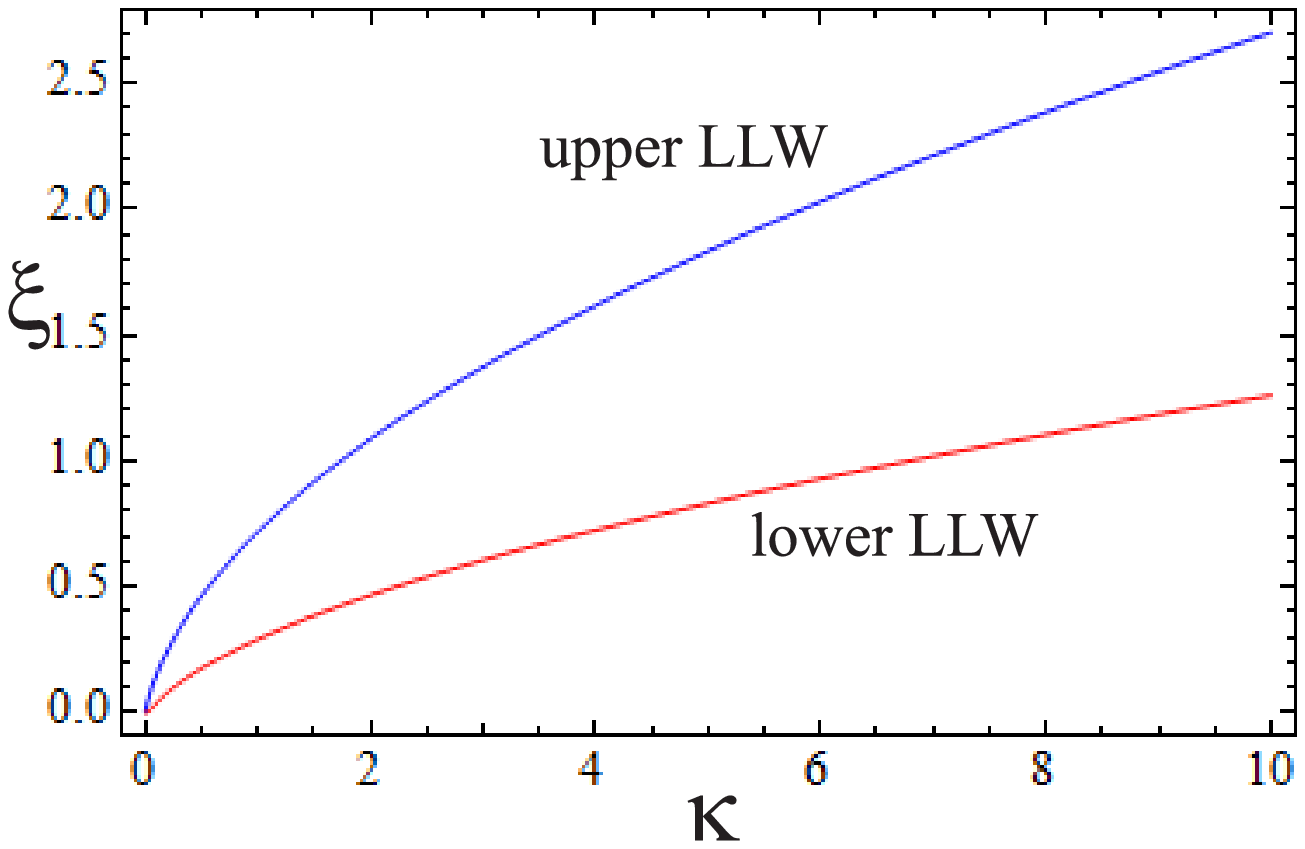}
\caption{\label{SUSD ObEx 05}
The high-frequency solutions of equation (\ref{SUSD_2L2D GENERAL spectrum}) are demonstrated at the following parameters
$c_{1}=0.2$, $\eta_{1}= 0.1$, $c_{2}=0.8$, $\eta_{2}=0.2$, $b=0.2$, $r=0.0009$, $q = 10$ (no quantum Bohm potential is included).}
\end{figure}

\begin{figure}
\includegraphics[width=8cm,angle=0]{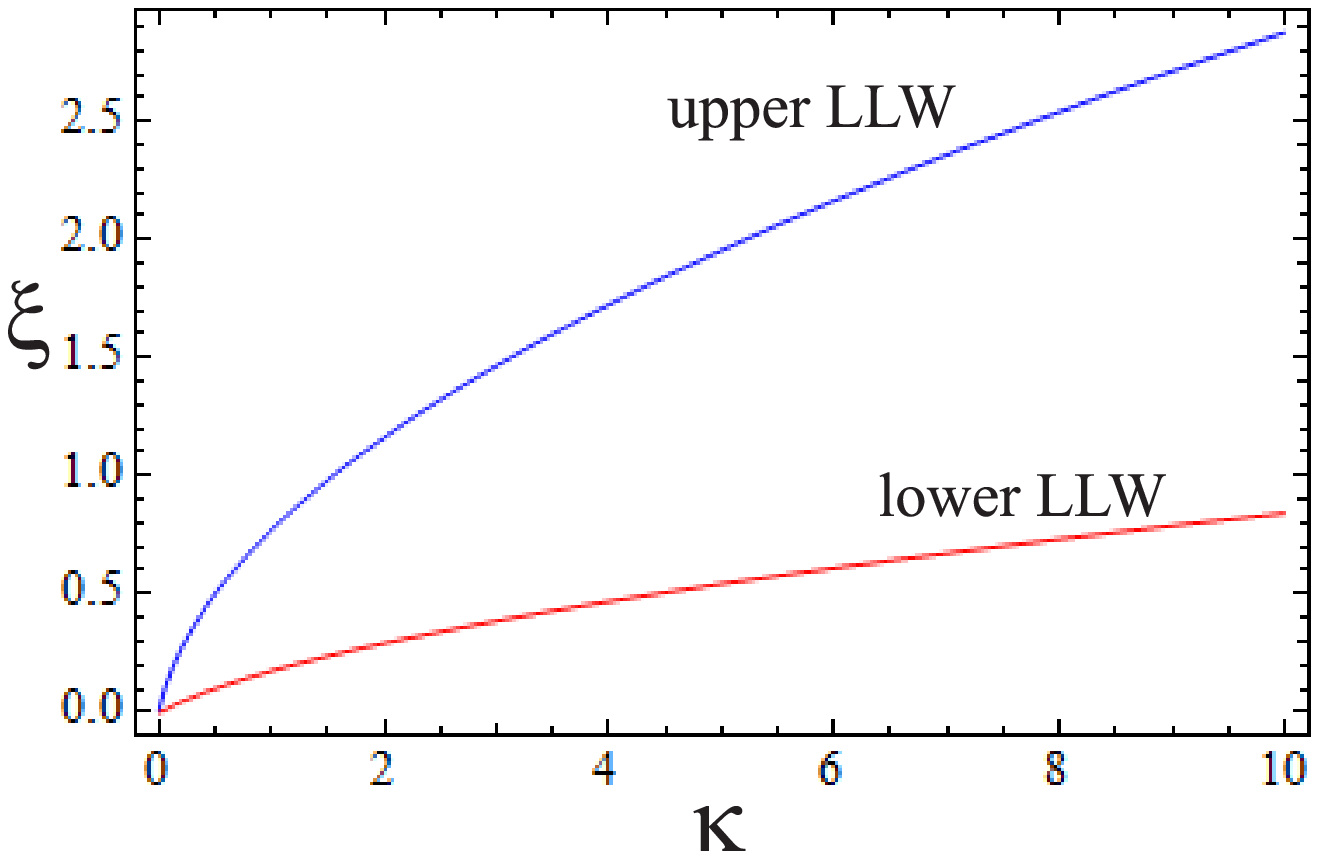}
\caption{\label{SUSD ObEx 06}
The high-frequency solutions of equation (\ref{SUSD_2L2D GENERAL spectrum}) are demonstrated at the following parameters
$c_{1}=0.1$, $\eta_{1}= 0.1$, $c_{2}=0.9$, $\eta_{2}=0.2$, $b=0.2$, $r=0.0009$, $q = 10$ (no quantum Bohm potential is included).}
\end{figure}

\begin{figure}
\includegraphics[width=8cm,angle=0]{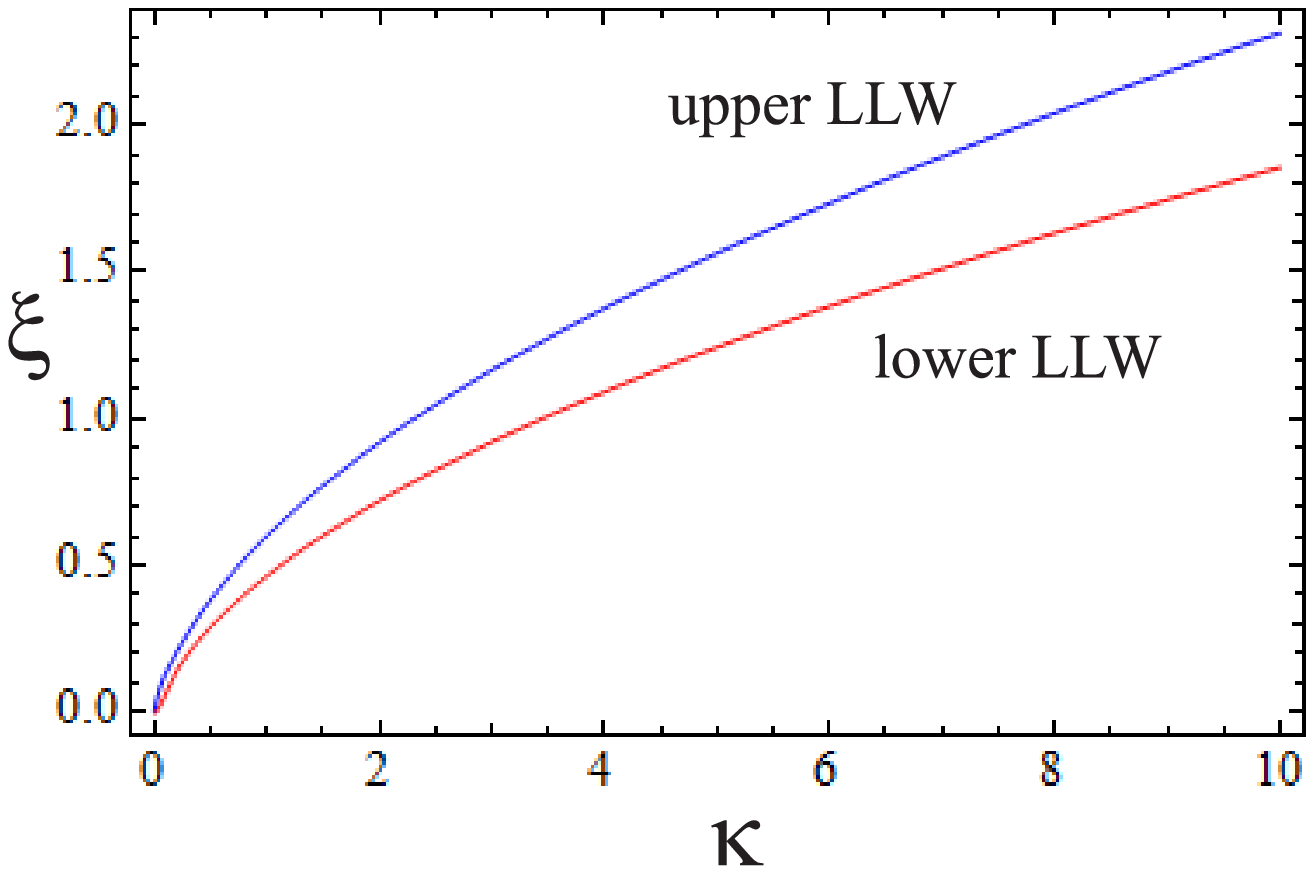}
\caption{\label{SUSD ObEx 07}
The high-frequency solutions of equation (\ref{SUSD_2L2D GENERAL spectrum}) are demonstrated at the following parameters
$c_{1}=0.4$, $\eta_{1}= 0.1$, $c_{2}=0.6$, $\eta_{2}=0.2$, $b=0.2$, $r=0.0009$, $q = 10$ (no quantum Bohm potential is included).}
\end{figure}

\begin{figure}
\includegraphics[width=8cm,angle=0]{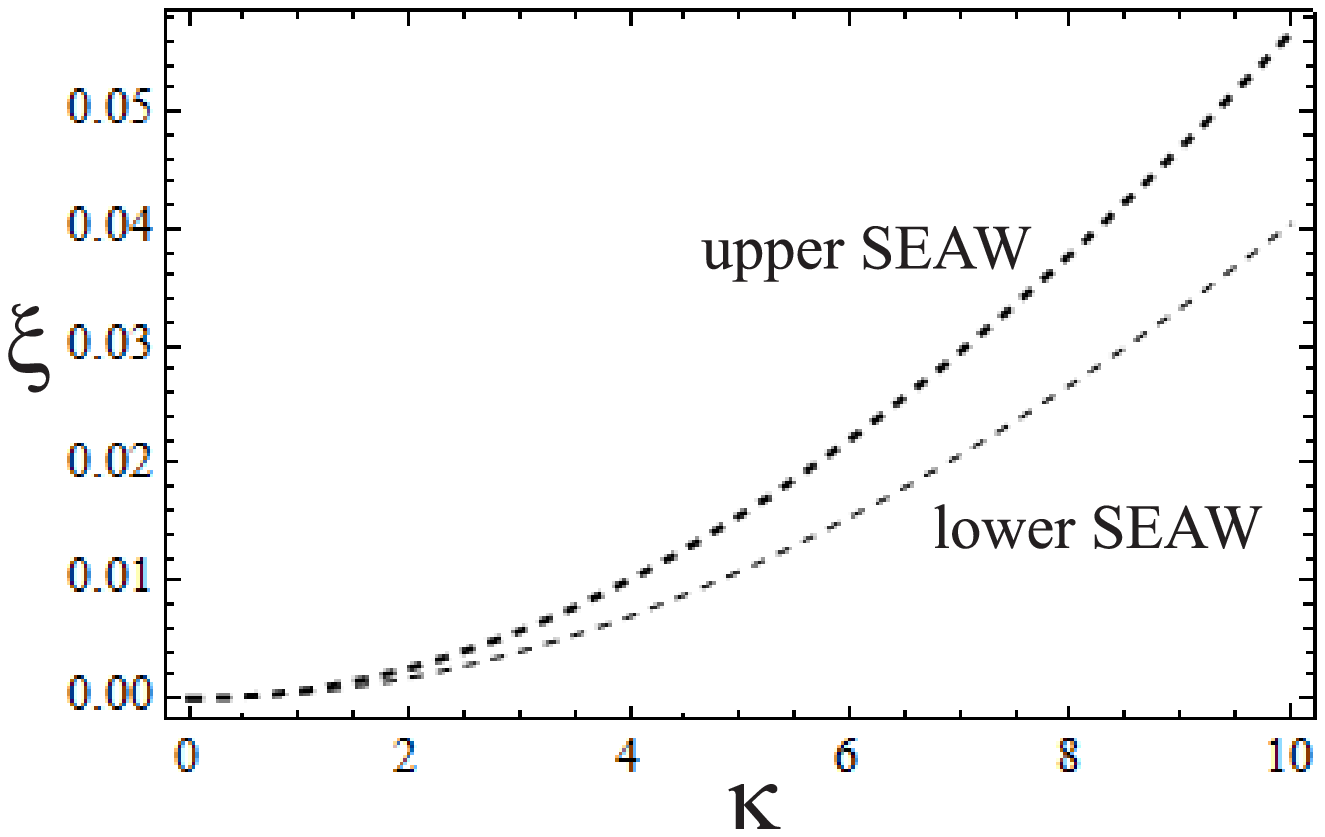}
\caption{\label{SUSD ObEx 08}
The low-frequency solutions of equation (\ref{SUSD_2L2D GENERAL spectrum}) are demonstrated at the following parameters
$c_{1}=0.4$, $\eta_{1}= 0.1$, $c_{2}=0.6$, $\eta_{2}=0.2$, $b=0.2$, $r=0.0009$, $q = 10$ (no quantum Bohm potential is included).}
\end{figure}

The last three terms existing in the Euler equations (\ref{SUSD2D Euler eq electrons spin UP}) describe the spin-spin interaction.
The first two of these terms presents the traditional force field of dipole-dipole interaction,
but written for electron subspecies.
The last term describes the spin-torque contribution related to the change of the spin direction,
which reveals in the particle number in the spin-s electron subspecies.
It exists in the Euler equation since the change the particle number leads to the change of the momentum of the subspecies.

The spin imbalance $\Delta n_{j}=n_{0ju}-n_{0jd}$ is caused  by external magnetic field.
Since electrons are negatively charged,
their spins get preferable direction opposite to the external magnetic field
$\eta_{j}\equiv\frac{\Delta n_{j}}{n_{0ej}}=\tanh(\gamma_{e}B_{0}/\varepsilon_{Fj})$,
where $\varepsilon_{Fj}=\pi n_{0ej}\hbar^{2}/m$ is the Fermi energy of the 2DEG, and $n_{0ej}=n_{0ju}+n_{0jd}$.
The system can be characterized by the fool spin polarization.
However, we consider dynamic of each constructive elements.
Therefore, we use the partial spin polarizations and the partial spin densities either.
The work is focused on the magnetically ordered materials.
Therefore, the spin-polarization of electrons are affected by the effective inner magnetic field
$\eta_{j}=-\tanh(\mid\gamma_{e}\mid (B_{0}+B_{eff,j})/\varepsilon_{Fj})$.
It is assumed that the multi-layered structures can be created as the combinations of different materials.
Hence, the effective magnetic field can be different in different 2DEGs $B_{eff,j}$
which is pointed out via subindex $j$ in the effective magnetic field.

The following comment about the quantum Bohm potential is essential.
The general many-particle form of the quantum Bohm potential in the Euler equation is found in Ref.
\cite{Maksimov QHM 99-01} (see equation 29).
Let us stress that it is found with no assumptions.
Same result is demonstrated in Ref. \cite{Andreev 1407.7770 Rev} by equation (23),
but it is written in another identical form.
However, this general form does not allow to solve any problem,
since it is not written via the hydrodynamic variables.
It is problematic to find a well justified approximate fermions even for the degenerate fermions.
The traditional formula is for the single particle \cite{Takabayasi PTP 55 a}
(it does not related to plasmas),
or for the case all particles in the same state
(it does not related to the fermions,
but for the bosons in the Bose-Einstein condensate state \cite{Andreev PRA08}).
Nevertheless,
if we consider the linear waves,
there is a general equation.
It is presented in the text after eq. (23) in Ref. \cite{Andreev 1407.7770 Rev} by the first of two terms.
Let us rewrite it here (with the missprint correction)
$-\frac{\hbar^2}{4m}\nabla\triangle n$.
It is the correct linear part for all kind of particles and geometries.
The nonlinear part requires some approximations.

\begin{figure}
\includegraphics[width=8cm,angle=0]{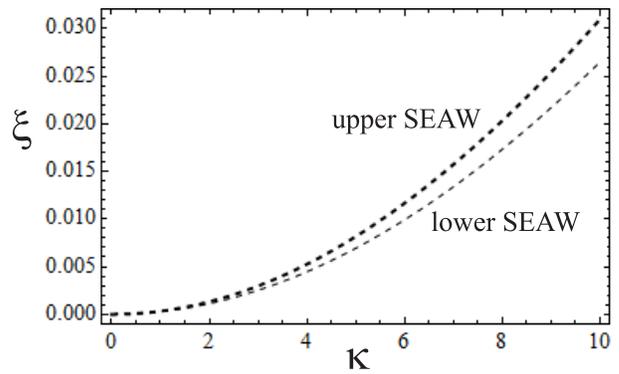}
\caption{\label{SUSD ObEx 09}
The low-frequency solutions of equation (\ref{SUSD_2L2D GENERAL spectrum}) are demonstrated at the following parameters
$c_{1}=0.3$, $\eta_{1}= 0.1$, $c_{2}=0.7$, $\eta_{2}=0.8$, $b=0.2$, $r=0.0009$, $q = 10$ (no quantum Bohm potential is included).}
\end{figure}

\begin{figure}
\includegraphics[width=8cm,angle=0]{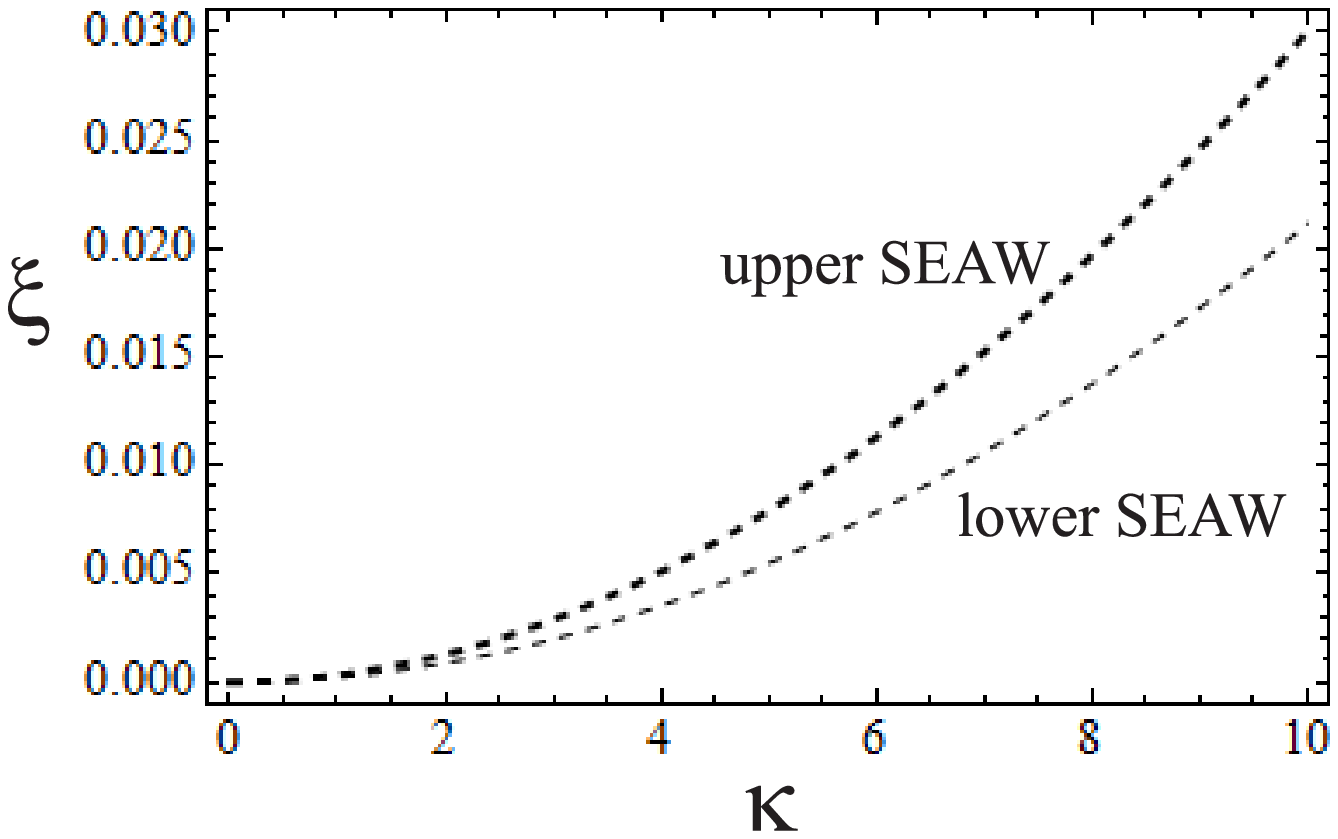}
\caption{\label{SUSD ObEx 10}
The low-frequency solutions of equation (\ref{SUSD_2L2D GENERAL spectrum}) are demonstrated at the following parameters
$c_{1}=0.2$, $\eta_{1}= 0.1$, $c_{2}=0.8$, $\eta_{2}=0.8$, $b=0.2$, $r=0.0009$, $q = 10$ (no quantum Bohm potential is included).}
\end{figure}

\begin{figure}
\includegraphics[width=8cm,angle=0]{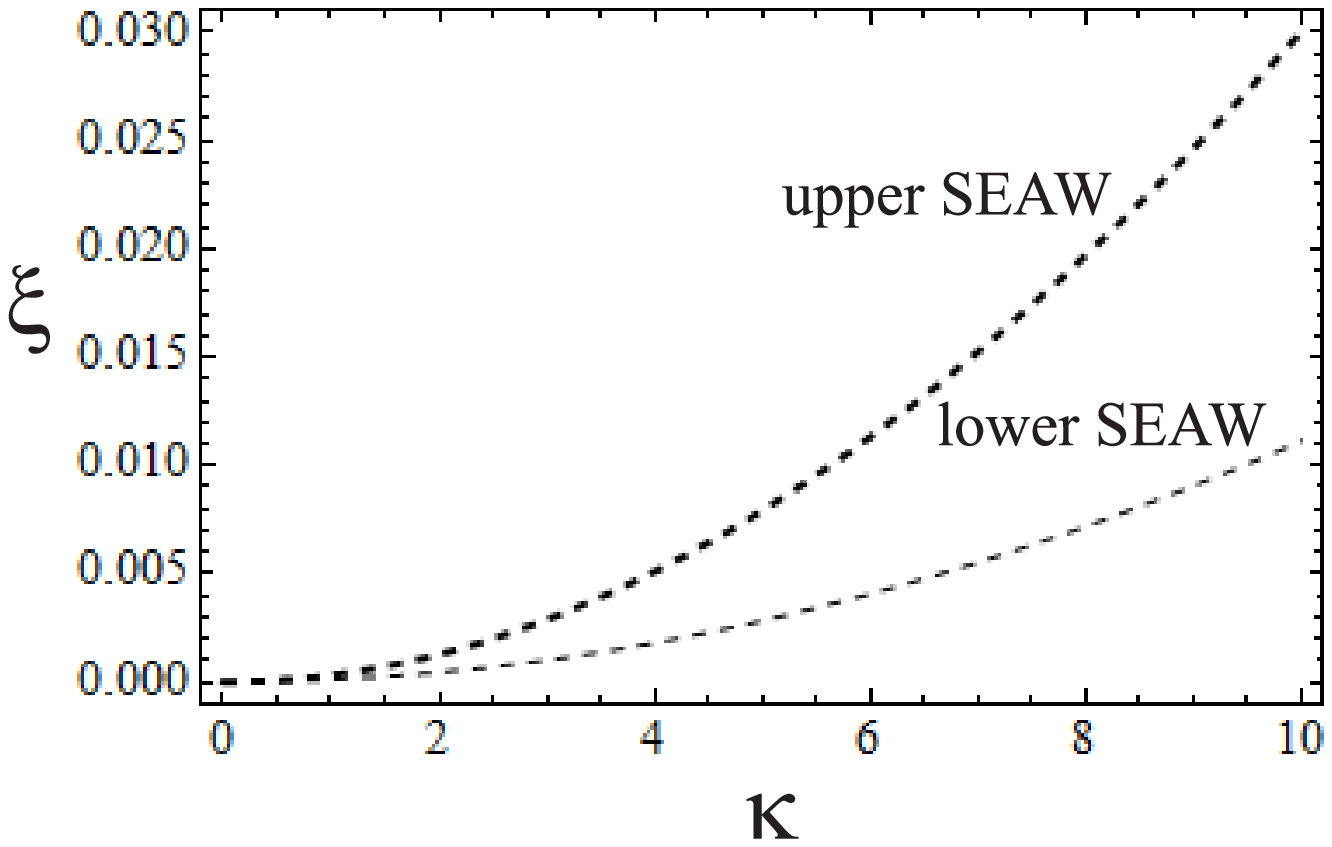}
\caption{\label{SUSD ObEx 11}
The low-frequency solutions of equation (\ref{SUSD_2L2D GENERAL spectrum}) are demonstrated at the following parameters
$c_{1}=0.2$, $\eta_{1}= 0.7$, $c_{2}=0.8$, $\eta_{2}=0.8$, $b=0.2$, $r=0.0009$, $q = 10$ (no quantum Bohm potential is included).}
\end{figure}

\begin{figure}
\includegraphics[width=8cm,angle=0]{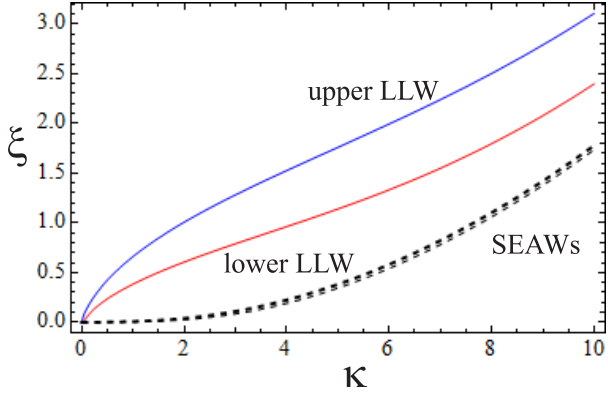}
\caption{\label{SUSD ObEx 12}
The solutions of equation (\ref{SUSD_2L2D GENERAL spectrum}) are demonstrated at the following parameters
$c_{1}=0.3$, $\eta_{1}= 0.1$, $c_{2}=0.7$, $\eta_{2}=0.2$, $b=0.2$, $r=0.009$, $q = 10$.
The influence of the quantum Bohm potential is demonstrated.}
\end{figure}

\begin{figure}
\includegraphics[width=8cm,angle=0]{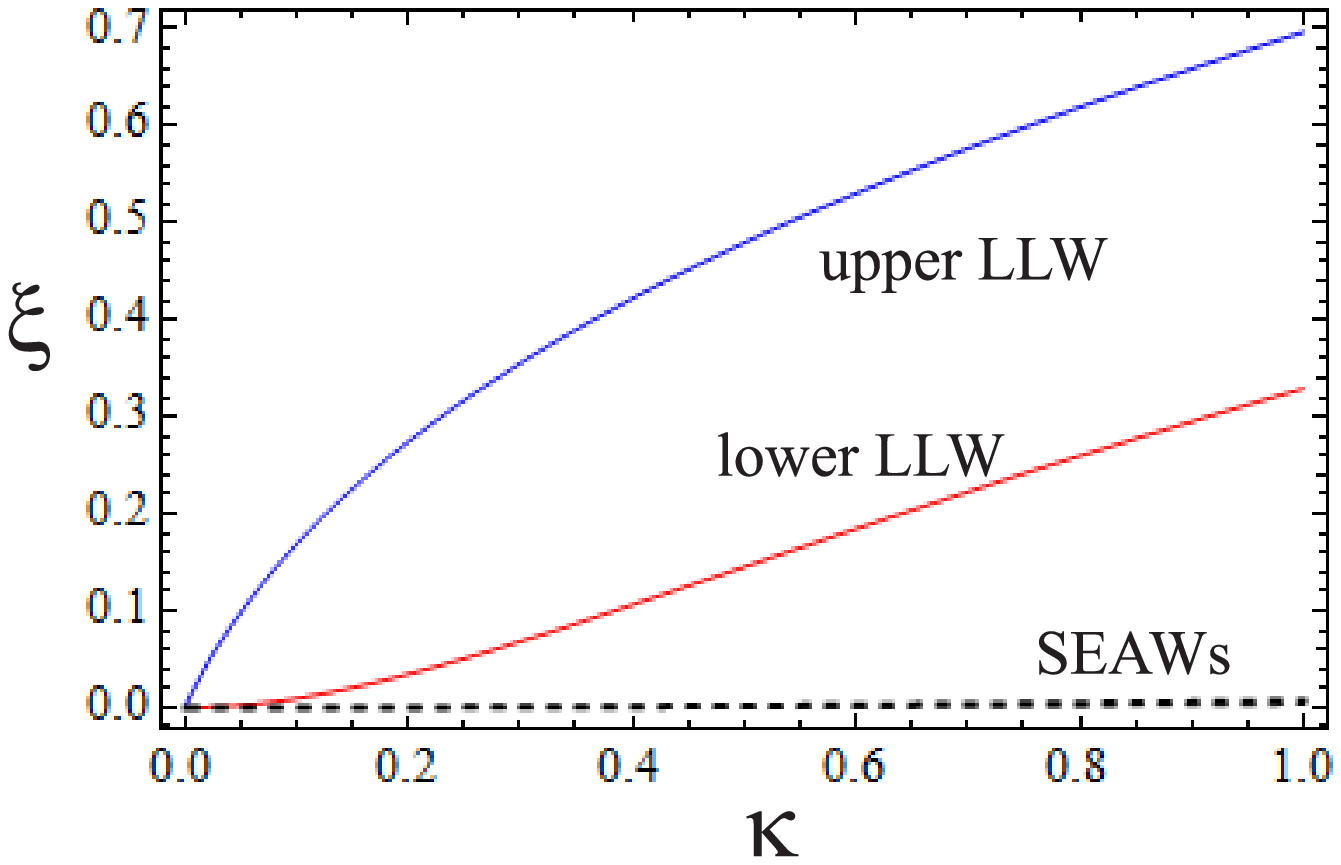}
\caption{\label{SUSD ObEx 13}
The solutions of equation (\ref{SUSD_2L2D GENERAL spectrum}) are demonstrated at the following parameters
$c_{1}=0.3$, $\eta_{1}= 0.1$, $c_{2}=0.7$, $\eta_{2}=0.2$, $b=0.2$, $r=0.009$, $q = 1$ (no quantum Bohm potential is included).}
\end{figure}

\begin{figure}
\includegraphics[width=8cm,angle=0]{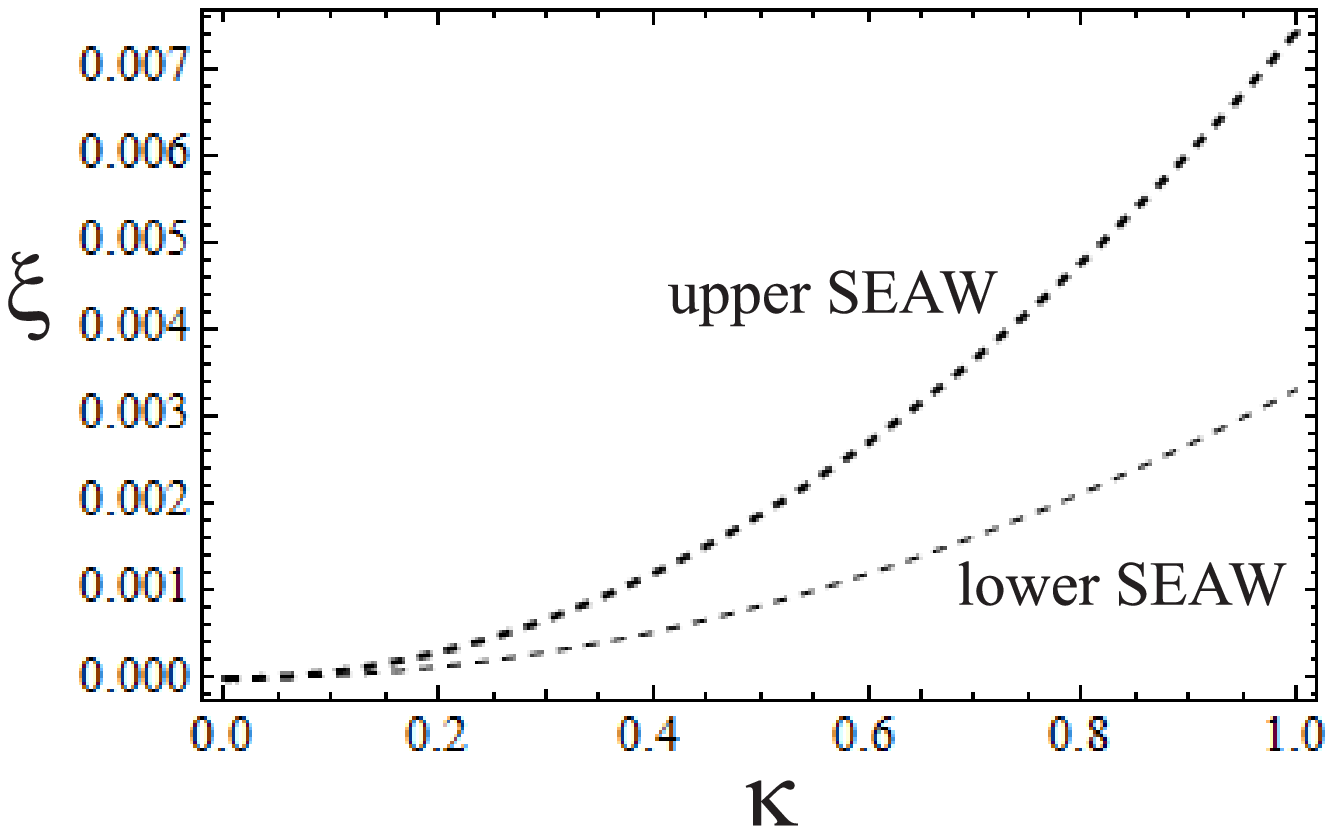}
\caption{\label{SUSD ObEx 14}
The low-frequency solutions of equation (\ref{SUSD_2L2D GENERAL spectrum}) are demonstrated at the following parameters
$c_{1}=0.3$, $\eta_{1}= 0.1$, $c_{2}=0.7$, $\eta_{2}=0.2$, $b=0.2$, $r=0.009$, $q = 1$ (no quantum Bohm potential is included).}
\end{figure}

\subsection{Wave solutions}

The SSE-QHD equations (\ref{SUSD2D cont eq electrons spin s})-(\ref{SUSD2D Euler eq electrons spin UP}) are rather complicate.
Hence, we present a linearized form of these equations for the electrostatic perturbations:
\begin{equation}\label{SUSD2D cont eq linearised}
-\imath\omega\delta n_{sj}
+\imath k_{x}n_{0sj}\delta v_{sj,x}=0, \end{equation}
$$n_{0sj}(-\imath\omega)\delta v_{sj,x}
+\frac{\imath k_{x}}{m} \frac{\partial p_{sj}}{\partial n_{sj}}\delta n_{sj}
+\frac{\hbar^{2}}{4m^{2}}k_{x}^{3}\delta n_{sj}=n_{0sj}v_{sj,y}\Omega$$
\begin{equation}\label{SUSD_2L2D Euler x linearised}
-\frac{\imath}{k} \omega_{Lsj}^{2}
(\delta n_{sj}+ e^{-kd}\delta n_{sj'}
+\delta n_{s'j}+ e^{-kd}\delta n_{s'j'}),\end{equation}
and
\begin{equation}\label{SUSD_2L2D Euler y linearised}
-\imath\omega\delta v_{sj,y}
=\Omega\delta v_{sj,x}. \end{equation}

Before the analysis of the SSE influence on the two layer system
we present SSE of electrons in the single layer of 2DEG.
We find spectrum consisting of two waves
which are the Langmuir wave and the SEAW (see Ref. \cite{Andreev EPL 16} eq. 7):
$$\omega^{2}-\Omega^{2}=\frac{1}{2}\Biggl(\omega_{Le}^{2}+(U_{u}^{2}+U_{d}^{2})k^{2}
\pm\biggl[\omega_{Le}^{4}$$
\begin{equation}\label{SUSD2D dispersion plane} +(U_{u}^{2}-U_{d}^{2})^{2}k^{4}
+2 k^{2}(U_{u}^{2}-U_{d}^{2})(\omega_{Lu}^{2}-\omega_{Ld}^{2})\biggr]^{1/2}\Biggr),\end{equation}
where $\omega_{Ls}^{2}=2\pi e^{2}n_{0s}k/m$ is the two dimensional Langmuir frequency for species $s$ of electrons located in the single layer, $\omega_{Le}^{2}=\omega_{Lu}^{2}+\omega_{Ld}^{2}$ is the full Langmuir frequency of the single layer,
$U_{s}^{2}=2\pi\hbar^{2}n_{0s}/m^2 +\hbar^{2}k^{2}/4m^2$,
$\Omega=q_{e}B_{0}/(mc)$ is the cyclotron frequency.

Two layers spectrum can be found from the following dispersion equation
which appears from equations (\ref{SUSD2D cont eq linearised})-(\ref{SUSD_2L2D Euler y linearised}):
$$[(\omega^{2}-\Omega^{2}-\omega_{L1u}^{2}-k^2 U_{1u}^2)(\omega^{2}-\Omega^{2}-\omega_{L1d}^{2}-k^2 U_{1d}^2) -\omega_{L1u}^{2}\omega_{L1d}^{2}]\times$$
$$\times[(\omega^{2}-\Omega^{2}-\omega_{L2u}^{2}-k^2 U_{2u}^2)(\omega^{2}-\Omega^{2}-\omega_{L2d}^{2}-k^2 U_{2d}^2) -\omega_{L2u}^{2}\omega_{L2d}^{2}]$$
$$-e^{-2kd}[(\omega^{2}-\Omega^{2})\omega_{L1e}^{2}-k^2(\omega_{L1u}^{2}U_{1d}^2+\omega_{L1d}^{2}U_{1u}^2)]\times$$
\begin{equation}\label{SUSD_2L2D GENERAL spectrum} \times[(\omega^{2}-\Omega^{2})\omega_{L2e}^{2}-k^2(\omega_{L2u}^{2}U_{2d}^2+\omega_{L2d}^{2}U_{2u}^2)]=0, \end{equation}
where $\omega_{Lsj}^{2}=2\pi e^{2}n_{0sj}k/m$ and $U_{sj}^{2}=2\pi\hbar^{2}n_{0sj}/m^2 +\hbar^{2}k^{2}/4m^2$.

Comparing equations (\ref{SUSD_2L2D classic spectrum}) and (\ref{SUSD_2L2D GENERAL spectrum})
we see that the SSE account changes number of waves existing in double layer system from two waves to the four waves.

The solutions of equation (\ref{SUSD_2L2D GENERAL spectrum}) are demonstrated in Figs. (\ref{SUSD ObEx 02})-(\ref{SUSD ObEx 14}).
It is found that two SEAWs appear in this regime.
The SEAWs have smaller frequencies then the Langmuir-like waves.
However, all dispersion curves starts at the cyclotron frequency $\mid\Omega\mid$ in the long-wavelength limit $k\rightarrow0$.
The frequencies of the Langmuir-like waves increase up to approximately $3\mid\Omega\mid$ and $4\mid\Omega\mid$
at the wave vector increase up to $\tilde{k}=10\sqrt{n_{0}}$ for the lover and higher LLWs, correspondingly.
While, the frequencies of the SEAWs increase up to $1.2\mid\Omega\mid$ and $1.4\mid\Omega\mid$
at the wave vector increase up to $\tilde{k}=10\sqrt{n_{0}}$
at $n_{01}+n_{02}\sim10^{12}$ cm$^{-2}$.
The group velocities of the SEAWs are considerably smaller.
The Langmuir-like waves have negative concavity (negative second derivative of the frequency on the wave vector $d^{2}\omega/dk^{2}$)
while the SEAWs have positive concavity.
The quantum Bohm potential changes the concavity of the lower LLW at the large wave vectors.


The influence of the quantum Bohm potential is considered separately.
It is presented in Fig. (\ref{SUSD ObEx 12}).
The quantum Bohm potential expectedly increases the frequencies of all waves being noticeable in the large wave vector regime $k\sim\tilde{k}=10\sqrt{n_{0}}$.


Dimensionless frequency of the Langmuir-like waves $\xi=\omega/\omega_{L0}$ does not depend on the concentration $n_{0}$ (change of parameter $r$)
and the spin polarizations $\eta_{j}$ (see Figs. (\ref{SUSD ObEx 02}) and (\ref{SUSD ObEx 03})).
There is a small increase of the frequency $\xi$ at the increase of $n_{0}$ at the large wave vectors $\kappa\sim10$,
but it is rather small.
However, it demonstrates a strong dependence on the difference of concentrations $n_{01}-n_{02}$.
It follows from comparison of Figs. (\ref{SUSD ObEx 03}), (\ref{SUSD ObEx 05}), (\ref{SUSD ObEx 06}), (\ref{SUSD ObEx 07}).
It is in general agreement with the approximate classic equation (\ref{SUSD_2L2D classic spectrum limit 1}).
Figs. (\ref{SUSD ObEx 03}), (\ref{SUSD ObEx 05}), (\ref{SUSD ObEx 06}) show that the upper (lower) LLW increases (decreases) its frequency at the increase of the concentration difference.
Corresponding decrease of the concentration difference (transition from Fig. (\ref{SUSD ObEx 03}) to Fig. (\ref{SUSD ObEx 07})) demonstrates the frequency decrease (increase) for the upper (lower) LLW.

Figs. (\ref{SUSD ObEx 04}), (\ref{SUSD ObEx 08}), (\ref{SUSD ObEx 09}), (\ref{SUSD ObEx 10}), (\ref{SUSD ObEx 11}) describe the spectrum of the SEAWs.
The SEAWs strongly depend on the concentration $n_{0}$ (see Figs. (\ref{SUSD ObEx 02}), (\ref{SUSD ObEx 03}), (\ref{SUSD ObEx 04}))
and the spin polarizations $\eta_{j}$ and on the difference of concentrations $n_{01}-n_{02}$.
The decrease of the frequency is demonstrated in Figs. (\ref{SUSD ObEx 02}), (\ref{SUSD ObEx 03}), (\ref{SUSD ObEx 04}) at the decrease of the concentration $n_{0}\sim r$.

The decrease of the concentration difference (the transition from Fig. (\ref{SUSD ObEx 04}) to Fig. (\ref{SUSD ObEx 08})) decreases (increases) the upper (lower) SEAW frequency.

Increase of the spin polarization of one of 2DEGs is presented by the transition from Fig. (\ref{SUSD ObEx 04}) to Fig. (\ref{SUSD ObEx 09}).
It shows the considerable decrease of frequencies of both SEAWs.
Moreover, the frequency difference of two SEAWs is decreases either.

Keeping the spin polarization like in Fig. (\ref{SUSD ObEx 09}) increase the difference of the concentrations and find Fig. (\ref{SUSD ObEx 10}).
In this case, the frequencies of both SEAWs decreases.
There is a small decrease for the upper SEAW while a noticeable decrease is found for the lower SEAW.
Hence, the character of the dependence on the concentration difference depends on the spin polarizations.

Next, increase the spin polarization of the second 2DEG in addition to the increased spin polarization of the first 2DEG and get the transition from Fig. (\ref{SUSD ObEx 10}) to Fig. (\ref{SUSD ObEx 11}).
It shows no changes in the spectrum of the upper SEAW,
but it decreases the frequency of the lower SEAW.
Here the lowest spin polarization is increased up to the value which is lowest again.
The transition of the value of the spin polarization of 2DEG with the lowest spin  polarization can be associated with the lowest SEAW
(propagating in this (first) 2DEG if the interaction between layers is neglected).

Comparison of Fig. (\ref{SUSD ObEx 02}) and Fig. (\ref{SUSD ObEx 12}) demonstrates the contribution of the quantum Bohm potential.
As it is mentioned above the frequency of all four waves increases due to the quantum Bohm potential contribution,
but low-frequency solutions are affected more strongly.

The distance between 2DEGs related to the intensity of the Coulomb interaction of 2DEGs.
Decrease the distance (transition from Fig. (\ref{SUSD ObEx 02}) to Figs (\ref{SUSD ObEx 13}) and (\ref{SUSD ObEx 14})) drops the frequencies of all waves.

\section{Conclusion}

Two layers of the plane-like 2DEGs separated by distance $d$ have been studied
in the regime of large potential barrier between two 2DEGs, hence no particle exchange has been assumed.
The hydrodynamic models have been applied for the theoretical description of waves in these systems.

The single 2DEG shows the Langmuir wave with the approximately square root spectrum $\omega\sim\sqrt{k}$.
The two interacting 2DEG demonstrates spectrum of two waves.
Their physical interpretation in the long-wavelength limit is the Langmuir-like waves with the interference-like dependence on concentrations with different signs for different wave.
It has been found that the SSE doubles the number of waves in the system.
The SEAW appears in the single 2DEG.
While, two SEAWs have been found in the double layer.

Properties of all four waves have been studied numerically.
In accordance with the single 2DEG,
it has been found that the frequency of the SEAWs is considerably smaller then the frequency of the LLWs.


\end{document}